# Towards Application of the RBNK Model

Larry Bull

University of the West of England

Bristol BS16 1QY, U.K.

Larry.bull@uwe.ac.uk

Abstract

The computational modeling of genetic regulatory networks is now common place – either by fitting a system to experimental data or by exploring the behaviour of abstract systems with the aim of identifying underlying principles. This paper presents an approach to the latter, considering the response to environmental changes of a well-known model placed upon tunable fitness landscapes. The effects on genome size and gene connectivity are explored.

# Introduction

With the aim of enabling the systematic exploration of artificial genetic regulatory network models (GRN), a simple approach to combining them with abstract fitness landscapes has recently been presented [Bull, 2012]. More specifically, random Boolean networks (RBN) [Kauffman, 1969] were combined with the NK model of fitness landscapes [Kauffman & Levin, 1987]. In the combined form – termed the RBNK model – a simple relationship between the states of $N$ randomly assigned nodes within an RBN is assumed such that their value is used within a given NK fitness landscape of trait dependencies.

In this paper, we tentatively begin to explore the use of the RBNK model to potentially capture some of the fundamental aspects of natural genomes, particularly their evolutionary responses to stress. Natural organisms experience many forms of stress and exhibit a multitude of responses with varying degrees of success in maintaining fitness. The underlying mechanisms and causes of such responses are of particular interest in areas beyond ecology, such as in plant and animal domestication. The effects of varying degrees of stress, cast as changes in the underlying fitness landscape, are explored with respect to genome size and gene connectivity in the RBNK model. The future aim is to compare the general results reported here with appropriate DNA analysis from crops and/or livestock, with some emphasis on the former envisaged here.

# Background

**The RBNK Model**

Within the traditional form of RBN, a network of $R$ nodes, each with a randomly assigned Boolean update function and $B$ directed connections randomly assigned from other nodes in the network, all update synchronously based upon the current state of those $B$ nodes. Hence those $B$ nodes are seen to have a regulatory effect upon the given node, specified by the given Boolean function attributed to it. Since they have a finite number of possible states and they are deterministic, such networks eventually fall into an attractor. It is well-established that the value of $B$ affects the emergent behaviour of RBN wherein attractors typically contain an increasing number of states with increasing $B$ (see [Kauffman, 1993] for an overview). Three phases of behaviour exist: ordered when $B=1$, with attractors consisting of one or a few states; chaotic when $B \geq 3$, with a very large number of states per attractor; and, a critical regime around $B=2$, where similar states lie on trajectories that tend to neither diverge nor converge (see [Derrida & Pomeau, 1986] for formal analysis). Figure 1 shows typical behaviour for various $B$. Such Boolean networks have been used within plant biology (e.g., [Greil, 2012]).

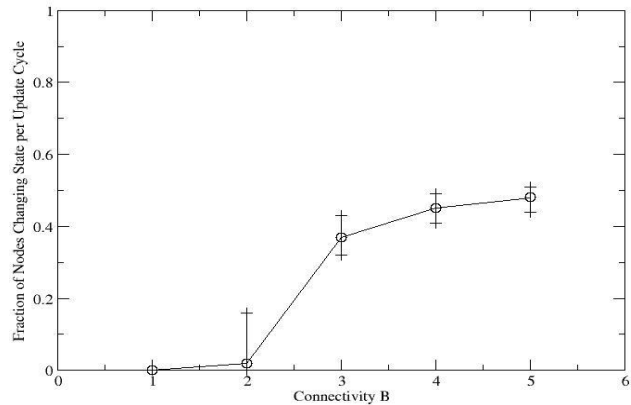

Figure 1. Typical behaviour of RBN with $R=100$ nodes and varying connectivity $B$, averaged after 100 update cycles over 100 runs. Nodes were initialized arbitrarily. Error bars show the min and max behaviour.

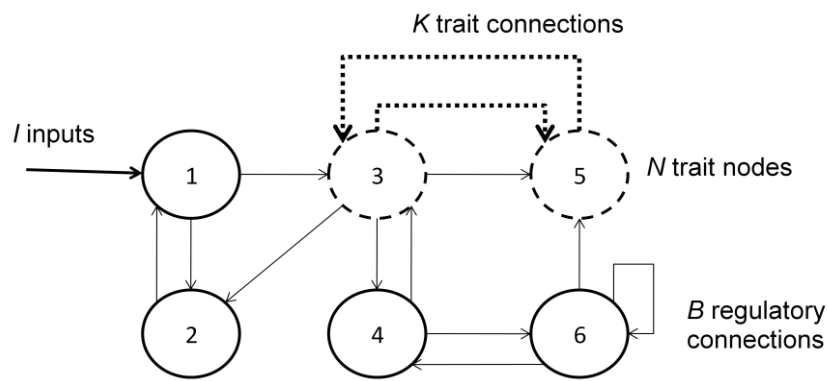

Figure 2. Example RBNK model with one input.

As shown in Figure 2, in the RBNK model *N* nodes in the RBN are chosen as "outputs", i.e., their state determines fitness using the NK model. Kauffman and Levin [1987] introduced the NK model to allow the systematic study of various aspects of fitness landscapes (see [Kauffman, 1993] for an overview). In the standard NK model an individual is represented by a set of *N* (binary) genes or traits, each of which depends upon its own value and that of *K* randomly chosen others in the individual. Thus increasing *K*, with respect to *N*, increases the epistasis. This increases the ruggedness of the fitness landscapes by increasing the number of fitness peaks.

The NK model assumes all epistatic interactions are so complex that it is only appropriate to assign (uniform) random values to their effects on fitness. Therefore for each of the possible *K* interactions, a table of $2^{(K+1)}$ fitnesses is created, with all entries in the range 0.0 to 1.0, such that there is one fitness value for each combination of traits. The fitness contribution of each trait is found from its individual table. These fitnesses are then summed and normalised by *N* to give the selective fitness of the individual. Exhaustive search of NK landscapes [Smith & Smith, 1999] suggests three general classes exist: unimodal when *K*=0; uncorrelated, multi-peaked when *K*>3; and, a critical regime around 0<*K*<4, where multiple peaks are correlated. The model has been used within plant biology (e.g., [Cooper & Podlich, 2002]).

The combination of the RBN and NK model enables a systematic exploration of the relationship between phenotypic traits and the genetic regulatory network by which they are produced. It was previously shown how achievable fitness decreases with increasing *B* and how increasing *N* with respect to *R* decreases achievable fitness [Bull, 2012]. In this paper *N* phenotypic traits are attributed to arbitrarily chosen nodes within the network of *R* genetic loci, with *I* environmental inputs applied to the first connection of the first *I* loci (Figure 2). Hence the NK element creates a tuneable component to the overall fitness landscape with behaviour (potentially) influenced by the environment.

**Simulations**

Following [Kauffman, 1993], the simple case of a greedy, genetic hillclimber is considered here. For a given offspring, mutation can either alter the size of the RBN or alter the Boolean function of a randomly chosen node or alter a randomly chosen connection for that node (equal probability). If the size is altered, one mutation deletes a randomly chosen node (the *N* trait nodes and *I* input nodes cannot be deleted), randomly re-assigning all of its connections, and one duplicates an existing node, connecting it to a randomly chosen node in the network. All four mutations happen with equal probability, one per offspring. A single fitness evaluation of a given RBN is ascertained by first assigning each node to a randomly chosen start state and updating each node synchronously for *T*=100 cycles. On each update cycle *T*, the value of each of the *N* trait nodes is then used to calculate fitness on the given NK landscape. This process is repeated ten times on the given NK landscape, the fitness assigned to the RBN being the average. Then a mutated RBN becomes the parent for the next generation if its fitness is higher than that of the original, or of it is smaller than the original, with ties broken at random. See [Bull, 2012] for an overview of previous work on evolving GRN.

Whilst the model is completely tunable, the simple scenario of a genome with expression data from ten genes is envisaged, i.e., *N*=10. Two further genes are imagined as each being sensitive to an environmental stimulus, i.e., *I*=2. Thus $R_{init}$=12 and following [Bull, 2012], results are averaged over 100 runs - 10 runs on each of 10 landscapes per parameter configuration - for 10,000 generations, with 0<*B*≤5 and 0≤*K*≤5 examined. Since there are two binary inputs, each of the four possible patterns is applied for *T*/4 cycles in sequence 00 to 11. Note the fitness landscape component remains the same during the application of all inputs. This is returned to below.

Figure 3 shows examples the main findings with this configuration of the RBNK model. As can be seen, there is only a significant difference (T-test, $p<0.05$) in fitness reached between *B*=1 and *B*>3, for low *K* (*K*<4). Moreover, such networks do grow slightly in size but to a significantly smaller amount (T-test, $p<0.05$) than for all other *B* in such cases. As the ruggedness of the fitness landscape increases, differences in fitness and size become less consistently significant.

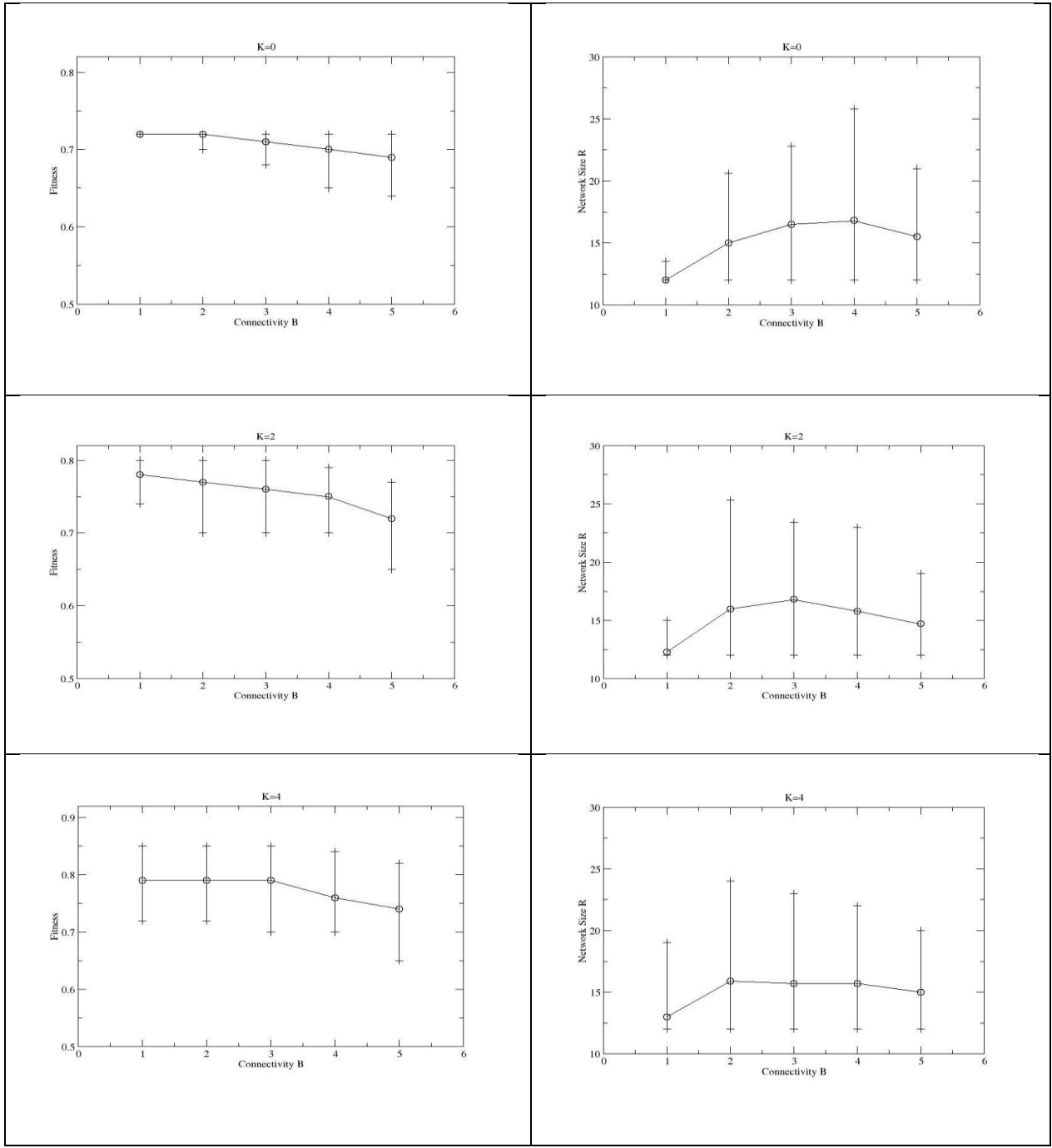

Figure 3. Evolutionary behaviour after 10,000 generations for the version of the RBNK model described in the text, for various *B* and *K* combinations.

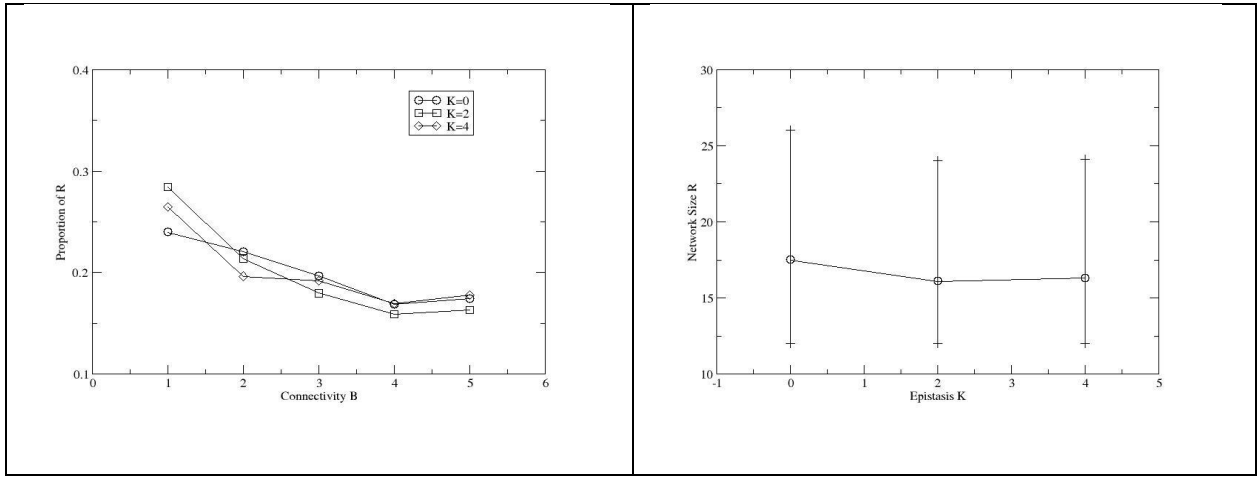

Figure 4. Evolutionary behaviour after 10,000 generations for the model when gene connectivity *B* is allowed to evolve, for various *K*.

It can be noted that within natural GRN genes have low connectivity on average (e.g., see [Leclerc, 2008]), as standard RBN can be seen to predict (Figure 1), but a number of high connectivity "hub" genes perform significant roles (e.g., see [Barabási & Oltvai, 2004]). As such, some previous work has explored allowing heterogeneity within RBN, i.e., allowing evolution to also define *B* for each node (e.g., see [Goudarzi et al., 2012]). That the average value of *B* tends towards the critical value of 2 in such cases is well-established for standard RBN (e.g., see [Kauffman, 1993]).

Figure 4 shows results for allowing mutation to also alter the value of *B* for a randomly chosen node within the range used above, i.e., [1, 5], with nodes being initialized randomly within the same range. As can be seen, networks evolve with *B*=1 as the most common degree of connectivity, although all other allowed values of *B* are seen at significant levels. The networks do not differ significantly (T-test, p≥0.05) in size for varying *K*. Fitness levels reached were not significantly different to those for low *B* in the fixed case above (not shown).

Aldana et al. [2007] have examined the effects of adding a new, single gene into a given RBN through a duplication mutation process. They find that the addition of one gene does not alter the attractor space of the resulting RBN when *B*=1 with highest probability. The probability of the transformed attractor landscape containing all the original attractors and at least a new one is maximised at *B*=2, but is also significant for *B*<4.

The results here suggest the (simple) evolutionary process errs on the side of caution for the input-landscape combinations tried as it makes more use of $B=1$ nodes than Aldana et al.'s findings anticipate.

As noted above, the NK model has previously been used with a view to plant biology in particular. Cooper and Podlich [2002] presented the E(NK) model in which $E$ environments are said to exist, each represented by an NK landscape which varies from the others by a tuneable degree (see [Bull, 1999] for a related study). Their aim was to begin to capture aspects of the degree of correlation between gene contributions to fitness and the correlation between, and frequency with which genomes experience, the different environments. Under various plant breeding-inspired selection strategies, they report $K$ has the most significant effect, which is perhaps unsurprising as the probability of the regions between global/high optima in the different landscapes being easily traversed will decrease with $K$; the number of local optima increases with $K$, as described above.

Following Cooper and Podlich [2002], the above RBNK models have been repeated such that $E=4$, that is, for each combination of input conditions, there exists a separate NK component (e.g., see [Bull, 2013] for a related study). Following [Bull, 1999], the fraction of the $N$ traits which differ between the landscapes is here termed $V$. Thus in both forms considered above, $V=0$. For $V>0$, the first $N \times V$ traits in each fitness landscape have their fitness table entries randomly re-assigned. Figure 5 shows examples of the effects of varying $V$ for the version of the model in which $B$ is also evolving, i.e., as in Figure 4. As can be seen, $R$ increases with $V$, such that the size of the networks when $V=1.0$ is significantly larger (T-test, $p<0.05$) than when $V=0.0$, for $K<5$. The same is also found to be true for fitness level, in reverse, i.e., it is significantly (T-test, $p<0.05$) decreased with increasing $V$ under the same conditions. There is no significant effect on the distributions of $B$ (not shown). The same has also been repeated for the fixed $B$ versions above and the same general result is found for $B<3$, with fitness always dropping significantly for all $B$ (not shown); as the variance in gene-by-environment increases, the underlying regulatory network increases in size.

This finding is related to that reported in [Crombach & Hogeweg, 2008] where simulated evolution fine-tunes regulatory network structure in response to an oscillatory non-stationary environment; a gene is added/removed under mutation with the addition/removal of the change in fitness landscape.

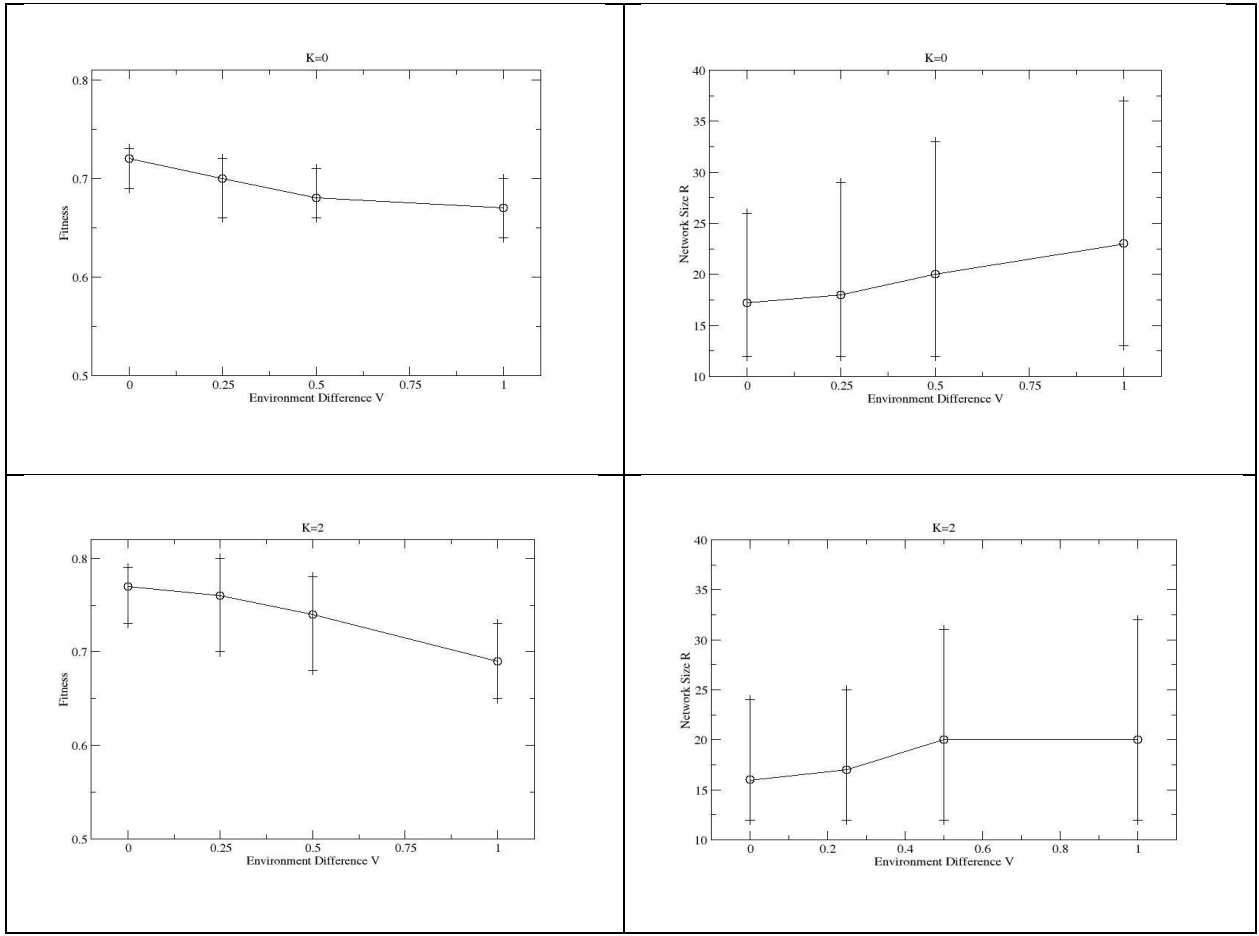

Figure 5. Evolutionary behaviour after 10,000 generations for the model when gene connectivity *B* is allowed to evolve, for various *K*, where a different fitness landscape is experienced for each of the four possible input combinations.

The robustness of these networks to lesser mutations has also been explored. In each test, the final GRN evolved was mutated by changing either a connection or node function, with 100 such one-mutant neighbours created. The type of mutation which produced the mutant with the lowest fitness was recorded. Results (not shown) indicate that a change in node function is roughly three times more likely to produce the lowest fitness mutant than altering a connection for *K*<5.

# Conclusions

Despite their simplicity, Boolean models of regulatory networks have been shown to have predictive capabilities (e.g., see [Shmulevich et al., 2002] for discussions and extensions). This paper has begun to explore the use of a variant of RBN to explore underlying phenomena in natural systems under stress. As noted above, the future aim is to align the approach to DNA data more closely across a variety of organisms. The area of motif formation within such models also remains to be explored in the near future.